\documentclass[twocolumn]{pasj00}
\usepackage{wrapfig}
\draft

\begin{document}
\title{Verification of the anecdote about Edwin Hubble and the Nobel Prize} 
\author{Kohji \textsc{Tsumura} \\
Frontier Research Institute for Interdisciplinary Science, Tohoku University, \\ 6-3 Aramaki-Aza-Aoba, Aoba-ku, Sendai, 980-8578, JAPAN}
\email{tsumura@astr.tohoku.ac.jp}
\maketitle
 
\begin{abstract}
Edwin Powel Hubble is regarded as one of the most important astronomers of 20th century.
In despite of his great contributions to the field of astronomy, he never received the Nobel Prize because astronomy was not considered as the field of the Nobel Prize in Physics at that era.
There is an anecdote about the relation between Hubble and the Nobel Prize. 
According to this anecdote, the Nobel Committee decided to award the Nobel Prize in Physics in 1953 to Hubble as the first Nobel laureate as an astronomer \citep{Christianson1995}.
However, Hubble was died just before its announcement, and the Nobel prize is not awarded posthumously.
Documents of the Nobel selection committee are open after 50 years, thus this anecdote can be verified.
I confirmed that the Nobel selection committee endorsed Frederik Zernike as the Nobel laureate in Physics in 1953 on September 15th, 1953, 
which is 13 days before the Hubble's death in September 28th, 1953.
I also confirmed that Hubble and Henry Norris Russell were nominated but they are not endorsed 
because the Committee concluded their astronomical works were not appropriate for the Nobel Prize in Physics.
\end{abstract}

\section{Introduction}
\subsection{Edwin Hubble}
Edwin Powel Hubble (1889-1953) is regarded as one of the most important astronomers of 20th century.
According to \citet{Sandage1989}, Hubble has four central accomplishments to the field of astronomy.
\begin{itemize}
 \item The galaxy classification system known as the Hubble morphological sequence of galaxy types.
 \item The discovery of Cepheids in NGC6822, with parallel work in M31 and M33, settling decisively the question of the nature of galaxies.
 \item The determination of the homogeneity of the distribution of galaxies, averaged over many solid angles.
 \item The linear velocity-distance relation known as the Hubble's Low.
\end{itemize}
He received the Barnard Medal in 1935, and all of 11 predecessors received the Nobel Prize.
However, he never received the Nobel Prize.
\ \\

\subsection{Astronomy on Nobel Prize}
The Nobel Prize is a set of annual international awards, and it is regarded as the highest honor in the field of the natural science.
However, astronomy is not in the list of natural sciences aimed at by the Nobel awards until 1960s.
Historically, Hans Bethe received the Nobel Prize in Physics in 1967 
{\it "for his contributions to the theory of nuclear reactions, especially his discoveries concerning the energy production in stars".}
Although his work is related to astronomy,  it is more like astrophysics.
The first Nobel Prize in Physics to the pure astronomical accomplishments is awarded to Mertin Ryle and Antony Hewish in 1974
{\it "for their pioneering research in radio astrophysics: Ryle for his observations and inventions, in particular of the aperture synthesis technique, and Hewish for his decisive role in the discovery of pulsars."}
\ \\

\subsection{An anecdote}
There is a famous(?) anecdote about Edwin Hubble and the Nobel prize.
According to the anocdote;
\begin{itemize}
 \item The Nobel Committee decided to award the Nobel Prize in Physics in 1953 to Hubble. 
 \item This would be the first Nobel prize to astronomy.
 \item However, Hubble was died just before its announcement, and the Nobel prize is not awarded posthumously.
\end{itemize}

This anecdote is often introduced in many reports and popular books of astronomy (e.g. \cite{Soraes2001, Singh2004, Iye2016}).
Probably, the origin of this anecdote is a biography, {\it "Edwin Hubble: Mariner of Nebulae"} \citep{Christianson1995}.
It says {\it "Grace\footnote{Edwin Hubble's wife} heard that Enrico Fermi and Subrahmanyan Chandrasekhar, both members of the Nobel Committee, 
had joined their colleagues in unanimously voting Hubble the prize in physics, a rumor later confirmed by the astronomers Geoffrey and Margaret Burbidge after speaking with Chandra."}

However, I got suspicious whether such a leak was really occurred from the members of the Nobel Committee,
because the Nobel prize is famous for its secrecy in the nomination process.
In this paper, I report the result of an investigation of the truth about this anecdote.
\ \\

\section{Verification}
\subsection{NobelPrize.org}
\begin{wrapfigure}{r}{80mm}
  \begin{center}
   \vskip -18mm
  \includegraphics[width=80mm]{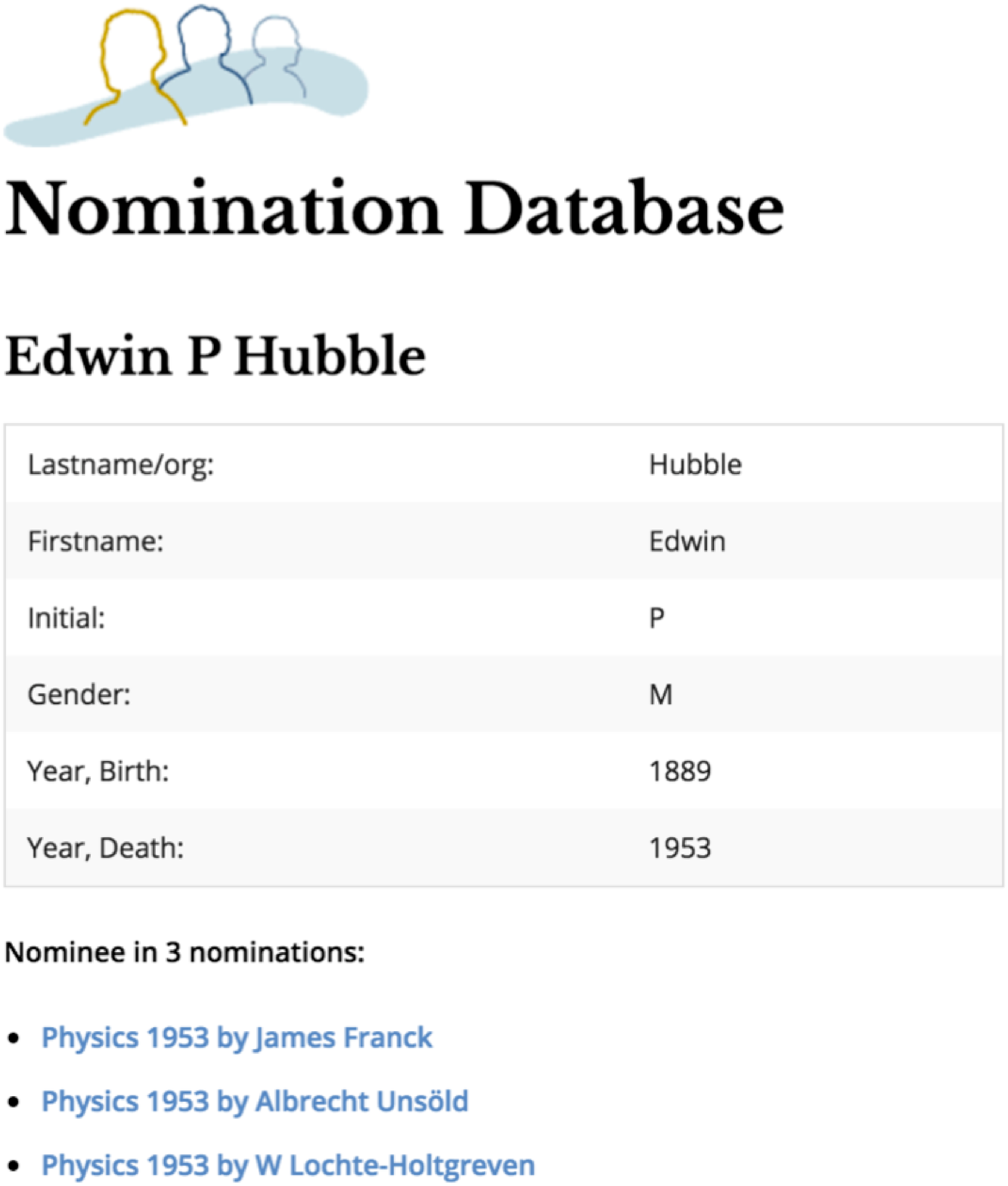}
    \end{center}
    \vskip -3mm
   \caption{Nomination of Edwin Hubble from NobelPrize.org database. \\ (http://www.nobelprize.org/nomination/archive/ show\_people.php?id=10529).}
  \label{nomination}
  \vskip -10mm
\end{wrapfigure}

First, I checked the official website of the Nobel Prize, NobelPrize.org, where all nominees and their nominators can be surveyed.
As a result, it is confirmed that Hubble was surely nominated in 1953 (the year of his death) by three nominators; James Franck, Albrecht Uns\"old, and W. Lochte-Holtgreven (Fig.\ref{nomination}).
However, decision process cannot be found from the database in the NobelPrize.org.

\subsection{Official Documents}
Official documents about decision process for the Nobel prize become open after 50 years, 
thus the document of selection process in 1953 is available from 2003.
Only the registered professionals can access the official documents.
\citet{Okamoto2016} checked the official documents about the selection process of Nobel Prize in Physics in 1953, and following facts were confirmed.
\begin{itemize}
   \item Edwin Hubble and Henry Norris Russell were nominated but they are not endorsed because the Committee concluded that their astronomical works were not appropriate for the Nobel Prize in Physics.
   \item Nobel selection committee endorsed Frederik Zernike as the Nobel laureate in Physics on September 15th, 1953.
   \item This decision was made 13 days before the Hubble's death in September 28th, 1953.
\end{itemize}
As a result, it is confirmed by checking the primary source materials that the well-known anecdote about the Edwin Hubble and the Nobel Prize is not true.

\section{Conclusion}
There is an anecdote that Edwin Hubble was selected for the Nobel Prize in Physics in 1953 but he was not awarded because he died before the announcement of it.
\citet{Christianson1995} says this rumor was {\it confirmed}, and many popular books about astronomy introduce this anecdote by referring it.
Investigation of the primary source materials, however, revealed this anecdote is not true, 
and it is confirmed that Hubble was not selected for the Nobel Prize in Physics in 1953, and this decision was made 13 days before he died.
It still unresolved issue why and how \citet{Christianson1995} {\it confirmed} this anecdote, but anyway, 
it is important to prevent the untrue rumor from spreading by many popular books.

\end{document}